\documentclass[preprint,12pt]{elsarticle}
\usepackage{amssymb}
\usepackage{multirow}
\usepackage{graphicx}
\usepackage{subfigure}
\usepackage{epsfig}
\usepackage[figuresright]{rotating}
\usepackage[dvips]{color}
\journal{Chemical Physics Letters}

\begin{document}
\begin{frontmatter}
\title{Structural, elastic, optical properties and quasiparticle band structure of solid cyanuric triazide}
\author{S. Appalakondaiah${^1}$ and G. Vaitheeswaran$^{1,*}$ and S. Leb\`egue${^2}$}
\address{1. Advanced Centre of Research in High Energy Materials (ACRHEM),
University of Hyderabad, Prof. C. R. Rao Road, Gachibowli, Andhra Pradesh, Hyderabad- 500 046, India. \\
2. Laboratoire de Cristallographie, R\'esonance Magn\'etique et Mod\'elisations (CRM2, UMR CNRS 7036), Institut Jean Barriol, Universit\'e de Lorraine, BP 239, Boulevard des Aiguillettes, 54506 Vandoeuvre-l\`es-Nancy, France. \\
\vspace{0.2in}
*Corresponding author E-mail address: gvsp@uohyd.ernet.in\\
\hspace {1.2in} Tel No.: +91-40-23138709\\
\hspace {1.2in} Fax No.: +91-40-23010227}
\end{frontmatter}
\newpage

\section*{Abstract}
In this letter, we report the structural, elastic, and quasiparticle band structure of cynauric triazide. The structural properties using a dispersion corrected method to treat
 van der Waals (vdW) forces offers a significant improvement in the description of the ground state properties. The predicted bulk modulus from the equation of
 state and the elastic constants are consistent and the magnitude lies in the order of secondary explosives. Then, the G$_0$W$_0$ approximation is used to study the band structure and an
 indirect band gap of 6.33 eV is obtained. Finally, we have calculated the optical and detonation characteristics at ambient pressure.
\newpage

\section{Introduction}
The azide (N$_3$) group combined with organic or inorganic ions exhibit great interest in the class of high energy materials due to their high heat of formation and release of abundant
 chemical energy in their decomposition process. For example, the  introduction of single azide into any functional group increases its energy by approximately 295-365 kJ/mol, which
 plays an important role to develop nitrogen rich materials\cite{book1}. These compounds are also used as good precursors to prepare bulk carbon nitride materials\cite{Cantillo,Miller,Ye}.
 The other major advantage of these azide based energetic compounds are towards producing molecular and non molecular form of nitrogen under pressure, which possess
 potential usage as green energetic materials\cite{Eremets1,Eremets2}.  Nevertheless, all these materials cannot be established in practical applications due to their extreme
 sensitivity with  external stimuli and their usage requires extreme precautions.  Hence, there is a particular attention in the synthesis and  design of these materials at the
 experimental and theoretical levels.
\par The presently studied compound (cyanuric triazide, C$_3$N$_3$(N$_3$)$_3$) is an organic azide, which belongs to the category of primary explosive. It is also used as a precursor to produce graphite particles and
 carbon nanotubes\cite{Kroke,Gillan}. The  presence  of carbon and  azide groups alone in this compound is beneficial to decompose as bulk
 carbon-nitride materials (C$_3$N$_4$ or C$_3$N$_5$) at extreme conditions\cite{Gillan}. Earlier, Hughes et al. reported the experimental crystal structure of  C$_3$N$_3$(N$_3$)$_3$ to be hexagonal
 with the space group H6$_3$/{\it m}\cite{Hughes}. Later, the Klap\"{o}tke group re-examined the crystal structure\cite{Elmer} and they found a structure different from the former one.
 From this study,  they concluded that C$_3$N$_3$(N$_3$)$_3$ crystalizes in an hexagonal structure with a different space group (P\={3}), with two molecules in the unit cell.
 It is a planar structure, containing the azide group with carbon and has a three-fold axis symmetry consisting of alternating carbon and nitrogen atoms in ring and the experimental crystal structure is shown in Fig. 1.
 On the theoretical side, by using ab-initio molecular dynamics simulations, Hu et al. predicted that the ground state structure will transform to nitrogen rich C$_3$N$_{12}$ at high pressures
 and temperatures\cite{Hu}. Apart from these, no other experimental or theoretical reports are available to explain the physical or chemical properties of C$_3$N$_3$(N$_3$)$_3$. Hence
 the opportunity is taken here to calculate the ground state properties with Density Functional Theory (DFT) and GW calculations. The content of the paper is as follows: A brief description of our computational details is presented in section 2. The results and discussion are presented in section 3 followed by summary of our conclusions in section 4.

\section{Computational details}
 The present calculations were performed using the plane wave pseudopotential method within  DFT as implemented in the CASTEP code\cite{Payne, Segall}.
 The structural optimization was done by the Broyden - Fletcher - Goldfarb - Shanno (BFGS) algorithm\cite{BFGS}. Vanderbilt-type ultrasoft pseudo potentials has been used
 to describe the electron-ion interactions\cite{Vanderbilt}. The exchange-correlation potential of Ceperley and Alder as parameterized by Perdew and Zunger (CAPZ) in the local density
 approximation (LDA)\cite{PPerdew} and also the generalized gradient approximation (GGA) with the Perdew-Burke-Ernzerhof (PBE)\cite{Perdew} parameterization were used to describe the
 electron-electron interactions. The convergence was reached when the forces on the atoms were less than 0.03 eV/\AA, the energies  were less than 1e-5 eV/atom
 and all the stress components were less than 0.05 GPa. To achieve this convergence, a plane wave basis set with an energy cutoff of 620 eV and a 5 $\times$ 5 $\times$ 7 Monkhorst-Pack
 mesh for the Brillouin zone sampling have been used\cite{Monkhorst}. In order to take into account van der Waals (vdW) interactions, we have used the semiemipirical dispersion correction as proposed
  by Grimme\cite{Grimme}. The total energy is therefore:
\begin{equation}
E_{total}=E_{DFT}+E_{Disp}
\end{equation}
where $E_{DFT}$ is the total energy of conventional DFT and the semi empirical dispersion correction is given by
\begin{equation}
E_{Disp}=-s_6\sum_{i<j}\frac{C_{ij}}{R^6_{ij}}f_{damp}(R_{ij})
\end{equation}
where C$_{ij}$ and R$_{ij}$ are the dispersion coefficient and interatomic distance between the i$^{th}$ and j$^{th}$ atoms, and s$_6$ is a global scaling factor.
A damping function, f$_{damp}$, is introduced to ensure that the dispersion correction is negligible at small R$_{ij}$. All the parameters taken from the original literature (except for s$_6$).

\par Also Kohn-Sham calculations within DFT usually underestimate band gaps and it is necessary to go beyond the standard exchange-correlation functionals. Hence, we have used another
 method to calculate the electronic gap  based on many body theory, which is the G$_0$W$_0$ approximation\cite{Hedin1,Hedin2}. The details of these calculations are presented in our previous papers\cite{konda1,konda2,seb1,seb2}. In the present
 case, we have used the G$_0$W$_0$ approximation as implemented in the VASP(Vienna Ab-initio Simulation Package)\cite{Kresse} code. To obtain convergence, we used
 200 bands for the summation over the bands in the polarizability and the self-energy formulas, and the polarizability matrices were calculated up to a cut-off of 150 eV.

\section{Results and discussion}
\subsection{Structural properties}
 Usually, vdW interactions appear to be major contributors to the binding forces within layered as well as molecular crystals, particularly in energetic materials. From earlier reports, standard DFT methods such
as local (LDA-CAPZ)  and semi-local (GGA-PBE) exchange and correlation functionals are inadequate to account for the non-local van der waals interactions in  these materials.
For this, we have used a correction proposed by Grimme\cite{Grimme}, which is found to work well for other energetic materials\cite{dftd,Landeville}. Hence, in the first step, we tested
 the accuracy of the present calculations by comparing the ground state structural properties with the experimental data. For this,
we started with the experimental crystal structure\cite{Gillan} of hexagonal C$_3$N$_3$(N$_3$)$_3$ as the starting geometry, then relaxed both the lattice parameters and atomic
 positions with convergence settings described in the previous section. Table I presents the calculated lattice parameters and volumes with standard DFT methods (LDA-CAPZ, GGA-PBE) and with the
 vdW corrected method (PBE+G06) along with experimental data. As compared with experiments, the volume obtained using the standard DFT methods is roughly
 lower by 9\% within LDA and higher by 35 \% with PBE. On the other hand, the calculated volume using the PBE+G06  method
is seen to provide a large improvement (overestimate by 2 \%) over the other methods relative to experiments. A similar trend has been observed for the {\it c}-axis (underestimate by ~8 \% with
 LDA, overestimate by ~9 \% with PBE,  and underestimate  ~2 \% with PBE+G06), which mainly involve the stacking layers binded through vdW interactions. Overall, PBE+G06 reproduces well the ground state
 structure, and therefore we will calculate the other properties with the PBE+G06 functional.
 \par Then we performed calculations to predict the equation of state (EOS) for C$_3$N$_3$(N$_3$)$_3$ crystal by fitting the pressure as a function of volume with the third order Birch-Murnaghan's
  equation of state. The pressure dependence of the volume and the lattice parameters up to 40 GPa is presented in Fig. 2. The volume curve clearly shows large compression in the low pressure region,
  i.e. below 12 GPa, and it can be seen that the unit cell volume decreases monotonically with $\Delta$V/V$_0$ = 28.5$\%$ from 0 to 12 GPa whereas it decreases to 17$\%$ from 16 to 40 GPa.
  In fact, in C$_3$N$_3$(N$_3$)$_3$ crystals,  the intermolecular distance along the {\it c}-axis is larger and therefore the interactions between the molecules are coming mostly from vdW forces.
  In general, at high pressures the long-range vdW forces become less effective,  which result in a high compressibility in the low pressure region. This can be clearly identified from
  the relative pressure coefficients ($\gamma = \frac{1}{X}\frac{dX}{dP}$, X={\it a},  {\it c}) of the lattice parameters. The calculated $\gamma$  values are for lattice parameters a and c
  are $\gamma$$_a$ = 8.9 $\times$ 10$^{-3}$, $\gamma$$_c$ = 7.1 $\times$ 10$^{-3}$ in the low pressure region (0 to 12 GPa) and these values decreases to $\gamma$$_a$ = 2.5 $\times$ 10$^{-3}$, $\gamma$$_c$ = 1.9
  $\times$ 10$^{-3}$ at high pressures ranging from 16 to 40 GPa respectively. At the same time, the compressibility  along two directions follow the sequence  c $<$ a, meaning that the compressibility
  of the present system is highly anisotropic. Then we fitted our results from 0 to 12 GPa to a Birch-Murnaghan EOS of C$_3$N$_3$(N$_3$)$_3$, and our calculated values of B$_0$ and
  B$_0$$^\prime$ are found to be 12.6 GPa and 6.42, respectively. The small value of B$_0$ indicate a very high compressibility of the material and the stiffness of the solid is increasing
  with pressure due to the large value of B$_0$$^\prime$.

\subsection{Elatic properties}
To understand the  mechanical stability and stiffness  of C$_3$N$_3$(N$_3$)$_3$,  we have performed calculations to predict its single crystal elastic constants.
To calculate the elastic properties,  we have used the Stress-Strain method.
  Since the present compound crystalizes in a hexagonal structure, it has five independent elastic constants: C$_{11}$,  C$_{33}$,  C$_{44}$, C$_{12}$ and
  C$_{13}$ that should satisfy Born's stability criteria\cite{born}:
C$_{ii}$$\textgreater$0(i=1,3,4), C$_{11}$$\textgreater$C$_{12}$ and  (C$_{11}$+C$_{12}$)C$_{33}$$\textgreater$2C$_{13}$$^2$ .  Our predicted single crystal elastic
 constants are presented in Table II and the calculated values satisfy the above mentioned criteria, which indicates that the system is mechanically stable at ambient pressure.
 In addition, we also found that $C_{11}$$>$$C_{33}$, which reveals considerable elastic anisotropic behavior in two crystallographic directions,  implies that {\it c}-axis
 is more compressible  than the {\it a}-axis and can easily undergo
 deformation under applied stress. Besides the single crystal elastic constants, we have also calculated the single crystal bulk modulus (B)
 of  C$_3$N$_3$(N$_3$)$_3$ from elastic constants. The obtained value is 10.45 GPa, which is slightly lower than the bulk modulus value obtained using Birch-Murnaghan EOS in the present work.
 Since, there are no experimental data available for the bulk modulus, we compare our values with the other energetic compounds.
  The experimental bulk moduli for inorganic azides are ranging from 16 to 40 GPa\cite{kondal}, on the other hand, the reported bulk moduli for conventional
  secondary explosives lies in between  8.3 to 20 GPa\cite{Landeville}. The computed bulk modulus for the present compound is in the order of secondary explosives, which might be an evidence that C$_3$N$_3$(N$_3$)$_3$ is a softer material than inorganic azides, and the compressibility is likely to be in the same order of well known known secondary explosives RDX, HMX, and PETN.

\subsection{Quasiparticle band structure}
 Zhu et al\cite{zhu} reported that band gaps of azides
 and explosives, and correlated them with their sensitivity as follows: the smaller the band gap, the easiest
 it is to transfer an electron from the valence bands to the conduction bands and consequently an external stimuli can more easily decompose
 the energetic material leading to explosion.
 Unfortunately, the experimental band gaps are not yet reported for most of these materials, the known band gaps values of inorganic azides (between 1.61 to 4.68 eV
 for heavy metal azides to light metal azides) and explosives materials (2.37 eV for TATB, 3.62 eV for HMX, 3.55 eV for RDX) are from standard DFT functionals.

 As stated above, standard DFT
 functionals have inherent limitation in reproducing experimental band gaps and advanced methods such as Hybrid functional like HSE\cite{Heyd,Krukau}, Tran and Blaha modified Becke
 Johnson potential\cite{Tran,david} or the GW approximation\cite{Mel,Jones} are available to overcome this problem. Hence, in the present work, we have calculated the
 electronic band structure using the experimental crystal structure of solid C$_3$N$_3$(N$_3$)$_3$, and the obtained G$_0$W$_0$ band structure along with the bandstructure
 obtained with the PBE functional is displayed in Figure. 3.
 From both bandstructures, it is found that an indirect band gap occurs between the M-point of the valence bands and the K-point of conduction bands. Moreover,  the
 magnitude of the band gap value increases from 3.65 eV to 6.33 eV from PBE to the GW approximation.  A similar increase in the band gap (PBE band gap to G$_0$W$_0$ band gap)
 values were obtained in our previous studies on secondary explosives such as solid nitromethane and FOX-7\cite{konda1,konda2}.

\par The electronic band structures along with  the partial density of states for various atoms (or groups) also provides clues on the nature of the bonding in the material.  We also calculated
the corresponding density of states (DOS) of C$_3$N$_3$(N$_3$)$_3$ with a scissor shift of 2.68 eV  in order to correct the band gap of PBE, the result being shown in Figure 4.
From the plot, the main orbital character of the bands can be described as follows: the conduction band maximum is from {\it p}- states of carbon, nitrogen
 (attached with carbon inside the ring) and azide group atoms.  At about -5 eV below, the valence band maximum was dominated by {\it s}- states of the azide group
 and {\it p}- states of all atoms. In the next region, around -3 eV, we find contribution of {\it s}- states of N atom, {\it p}- states of all C, N and azide group.
 At the Fermi level, the valence band maximum consists an overlap
 of {\it p}- states from azide group, nitrogen (inside the ring) and slight contribution of {\it s}- states from  nitrogen atom. These contributions determine
 the presence of  ionic bonding at the Fermi level, whereas covalent bonds from carbon to nitrogen are a dominant feature for the whole region.

\subsection{Optical properties}
Now, we focus on the optical properties of C$_3$N$_3$(N$_3$)$_3$. Optical spectroscopy is an important tool, mainly described by the dielectric tensor. The investigated compound
 being a molecular crystal and crystallizing in a hexagonal space group with high anisotropy from a structural point of view, it has only two nonzero components in the dielectric
 tensors: $\epsilon^{[100]} (\omega)$ =$\epsilon^{[010]} (\omega)$ and $\epsilon^{[001]} (\omega)$. In general, there are two contributions to $\epsilon (\omega)$  namely intraband and
 interband transitions: the contribution from intraband transitions is important only for the case of metals, and the interband transitions can be direct and indirect transitions. The indirect
 interband transitions involve scattering of phonons and give only a small contribution to $\epsilon (\omega)$ in comparison to the direct transitions, so we have neglected them in our
 calculations\cite{Smith}. Then $\varepsilon$(\emph{q=0}, $\omega$) was calculated using the random phase approximation without local field effects\cite{Cohen}. In particular, the imaginary
 part is determined by a summation over empty states, and then the real part $\epsilon_1 (\omega)$ of the dielectric function can be evaluated from $\epsilon_2 (\omega)$ using Kramer-Kroning
 relations. The knowledge of the real and imaginary parts of the dielectric function allows the calculation of the important optical properties such as refractive index and absorption. We have
 calculated the optical properties using the CASTEP code within GGA, a scissor operator correction of 2.68 eV has been applied in order to correct the standard DFT band gap. The absorptive
 (imaginary) part $\epsilon_2 (\omega)$ and the dispersive (real) part $\epsilon_2 (\omega)$ of the complex dielectric function $\epsilon (\omega)$ as a function of the photon energy are shown
 in Figs. 5(a) and (b) respectively. $\epsilon_2 (\omega)$ has four prominent peaks along the [100] direction due to the interband transitions between valence and conduction bands.  A sharp
 peak around ~7 eV, and a peak around 14.4 eV arises from N (inside the ring) {\it s}-states to C {\it p}-states, whereas the peak at 8.6 eV and 12.8 eV originate from
 Azide {\it s}-states to C {\it p}-states. On the other side, along the [001] direction, interband transition intensities are weaker compared to the [100] direction. This implies a considerable
 anisotropy for the absorptive part of dielectric function along the two different crystallographic directions of C$_3$N$_3$(N$_3$)$_3$. The static real part of dielectric function
 $\epsilon_1 (\omega)$, along the two crystallographic directions are found to be 3.01 (along [100]), and 1.17 (along [001]). Also, we have estimated the absorption functions of
 C$_3$N$_3$(N$_3$)$_3$ at two crystallographic directions from $\epsilon (\omega)$ and shown in Figure. 5(c).
 The first absorption peak for the two directions is observed at 7.1 eV, but the
 absorption coefficient is found to be 3.3 $\times$ 10$^5$ cm$^{-1}$ along the [100] direction and 2.1 $\times$ 10$^4$ cm$^{-1}$ along the [001] directions respectively.
 Apart from this, a strong absorption peak at 8.7 eV  is observed for the [100] direction.
 Since, the absorption coefficient is in the order of 10$^5$ cm$^{-1}$, the present compound is unstable and may decompose
 under the action of light. Finally, the refractive index of a crystal is closely related to the electronic polarizability of ions and local fields inside the crystal. The static
 refractive index is shown in Fig. 5 (d) and the  values of n(0) along the two crystal directions were found to be 1.72 (along [100]) and 1.08 (along [001]). From all the calculated
 optical constants, it is clearly observed that C$_3$N$_3$(N$_3$)$_3$ is an optically anisotropic material and is more sensitive to light (or electric field) along the [100] direction.

\subsection{Detonation properties}
Finally, we have also calculated important features for an energetic material such as the detonation velocity (D) and the detonation pressure (P) by the Kamlet-Jacobs equations\cite{Kamlet1,Kamlet2}.
 These equations are mainly based on the density  and heat of formation (HOF) to predict the detonation performance, the corresponding values for D (in km/s) and P (in GPa) being shown in following equations:
\begin{equation}
%D = 1.01(NM$^{0.5}$Q$^{0.5}$)$^{0.5}$(1+1.30$\rho$$_0$) and P = 1.55$\rho$$_0$$^2$NM$^{0.5}$Q$^{0.5}$
D= 1.01(NM^{0.5}Q^{0.5})^{0.5}(1+1.30\rho_0),  P = 1.55 \rho_0^2 NM^{0.5}Q^{0.5}
\end{equation}
In the above equations, N is the number of moles of gaseous detonation products per gram of explosives, M is average molecular weights of gaseous products, Q is chemical energy of detonation (kJ/mol) defined
 as the difference of the HOFs between products and reactants, and $\rho$$_0$ is the density of explosive (g/cm$^3$). We have calculated the detonation velocity and pressure using the calculated density
 (1.697 g/cm3) using PBE+G06 and the reported HOF (914.62 kJ/mol),
and the calculated values are  7.31 km/s and 22.87 GPa, respectively. Here, the calculated detonation velocity is higher than the earlier reports (5.5 km/s) \cite{pepekin} due to the
variations in the  density values,  which implies that the re-examined crystal structure by the Klap\"{o}tke group\cite{Elmer} shows prominent explosive properties. Apart from this, the calculated D
value exceeds the classical primary explosives and is in the order of conventional secondary explosives values such as RDX (8.5km/s),  PETN (8.4km/s), and TATB (7.3km/s)\cite{book2}.

\section{Conclusion}
In summary, the effect of van der Waals interactions on the ground  state structural properties of the layered molecular crystal C$_3$N$_3$(N$_3$)$_3$ was studied by dispersion corrected density functional theory.
 The structural properties calculated
through the van der Waals  correction to standard density functional method (PBE+G06) are in better agreement with the experimental results  compared with results obtained with LDA and GGA.
The obtained equilibrium crystal structure with PBE+G06 is further used to calculate the elastic constants. The calculated elastic constants follows an order of C$_{11}$$>$ C$_{33}$, indicating
that the material is more compressible along the c-axis. Including this, the obtained bulk modulus shows an high compressbility against mechanical stimuli. We have also found that the
 computed band gap is about 3.65 eV with GGA, and is enhanced to
6.33 eV with the G$_0$W$_0$ approximation, which reveals the importance of performing quasiparticle calculations for correlating the bandgap with the sensitivity in energetic materials . Also, we have observed
 a strong covalent bonding between carbon and nitrogen atoms from the density of states. The linear optical properties such as real and imaginary parts of the dielectric function,
refractive indices, and absorption spectra were also reported. The different possible electronic transitions and optical anisotropy  along different crystallographic directions were analyzed.
Finally, the detonation velocity and detonation pressure are calculated using Kamlet-Jacobs
equations from the experimental heat of formation and calculated density. C$_3$N$_3$(N$_3$)$_3$ shows explosive behavior, and the predicted detonation velocity and detonation pressure are 7.31 km/s and 22.87 GPa,
respectively. We hope that our work will  stimulate further experimental studies, in particular concerning the sensitivity and energetic properties of C$_3$N$_3$(N$_3$)$_3$.

\section{ACKNOWLEDGMENTS}
S. A. would like to thank DRDO through ACRHEM for financial support. S. A. and G. V. thank to DST through CMSD, University of Hyderabad, for providing computational facilities.
S. L. acknowledges the access to HPC resources from GENCI-CCRT/CINES (Grant x2014-085106).

\clearpage

\newpage
\textbf{Figure Legends:}  \\
Figure 1: (Colour Online) Experimental crystal structure of C$_3$N$_3$(N$_3$)$_3$  \\

Figure 2: (Colour online) The pressure dependence of lattice parameters and volume of C$_3$N$_3$(N$_3$)$_3$. The Equation of state fitted in the low pressure of C$_3$N$_3$(N$_3$)$_3$ were shown in inserted figures. \\

Figure 3: (Colour online) GGA (black lines) and G$_0$W$_0$ (red circles) band structure of C$_3$N$_3$(N$_3$)$_3$ at experimental lattice parameters.  \\

Figure 4: (Colour online) Total and partial density of states (DOS) of C$_3$N$_3$(N$_3$)$_3$ \\

Figure 5: (Colour online) Optical properties , (a) absorptive part of the dielectric function (b)  dispersive part of the dielectric function (c) refractive index and (d) absorption of C$_3$N$_3$(N$_3$)$_3$ as calculated at theoretical equilibrium with a scissor shift of 2.68 eV.\\

\clearpage

\begin{figure}
\begin{center}
{\includegraphics[width=6in,clip]{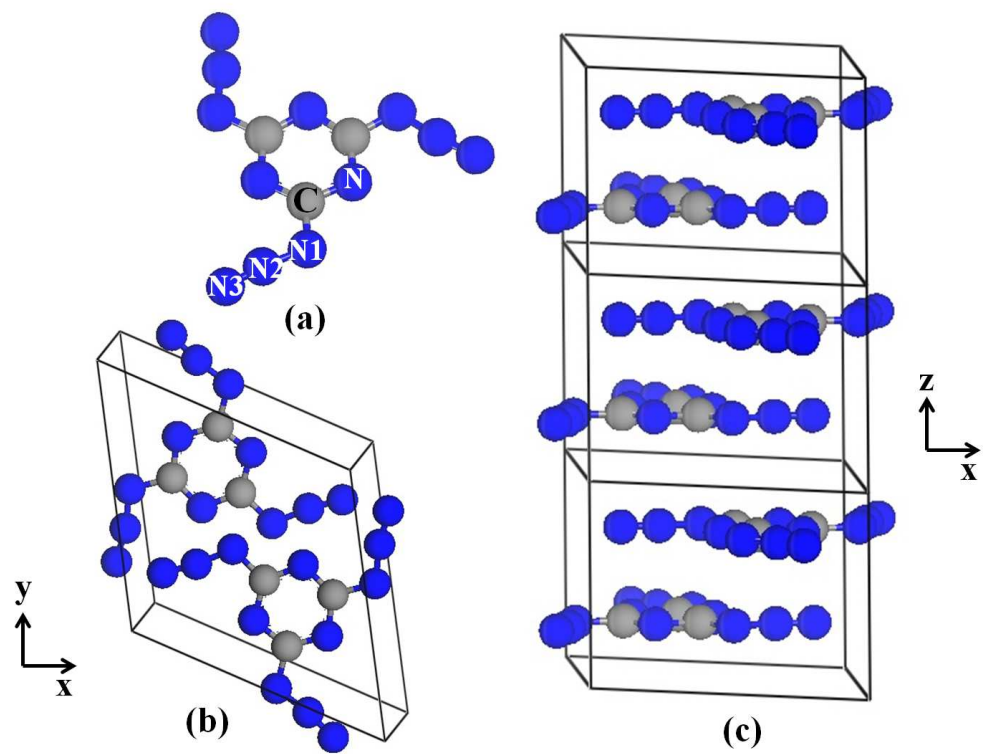}}
\caption{(Colour Online) Experimental crystal structure of C$_3$N$_3$(N$_3$)$_3$.}
\end{center}
\end{figure}
\clearpage
\newpage

\begin{figure}
\begin{center}
{\includegraphics[width=6.5in,clip]{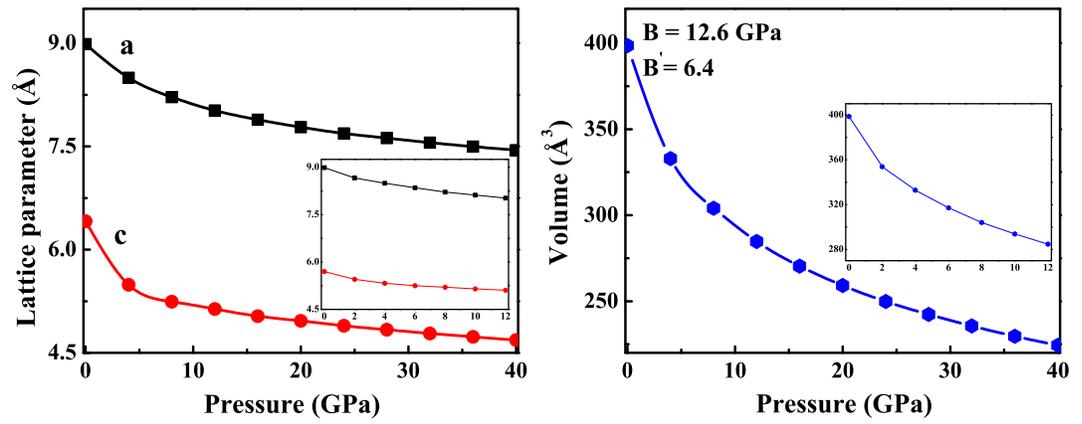}}
\caption{(Colour online) The pressure dependence of lattice parameters and volume of C$_3$N$_3$(N$_3$)$_3$. The Equation of state fitted in the low pressure of C$_3$N$_3$(N$_3$)$_3$ were shown in inserted figures.}
\end{center}
\end{figure}
\clearpage
\newpage

\begin{figure}
\begin{center}
{\includegraphics[width=5in,clip]{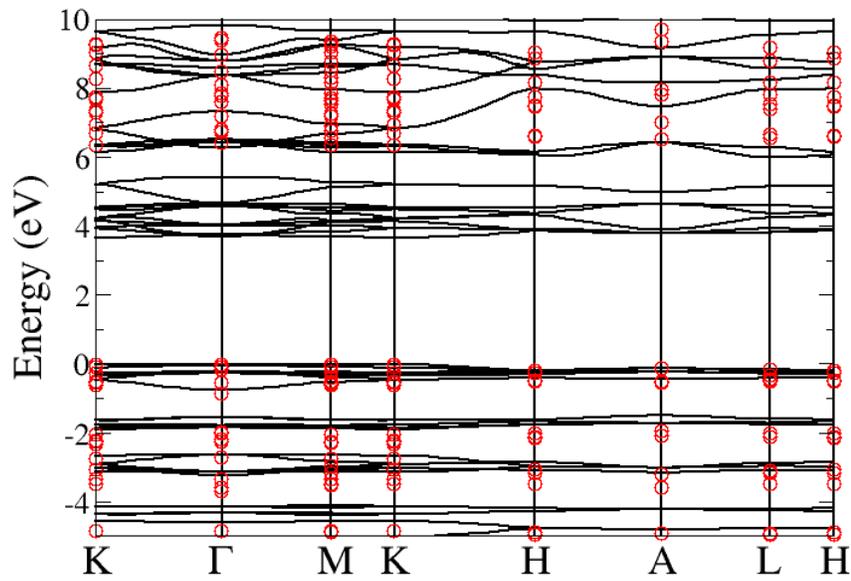}}
\caption{(Colour online) GGA (black lines) and G$_0$W$_0$ (red circles) band structure of C$_3$N$_3$(N$_3$)$_3$ at experimental lattice parameters.}
\end{center}
\end{figure}
\clearpage
\newpage

\begin{figure}
\begin{center}
{\includegraphics[width=5in,clip]{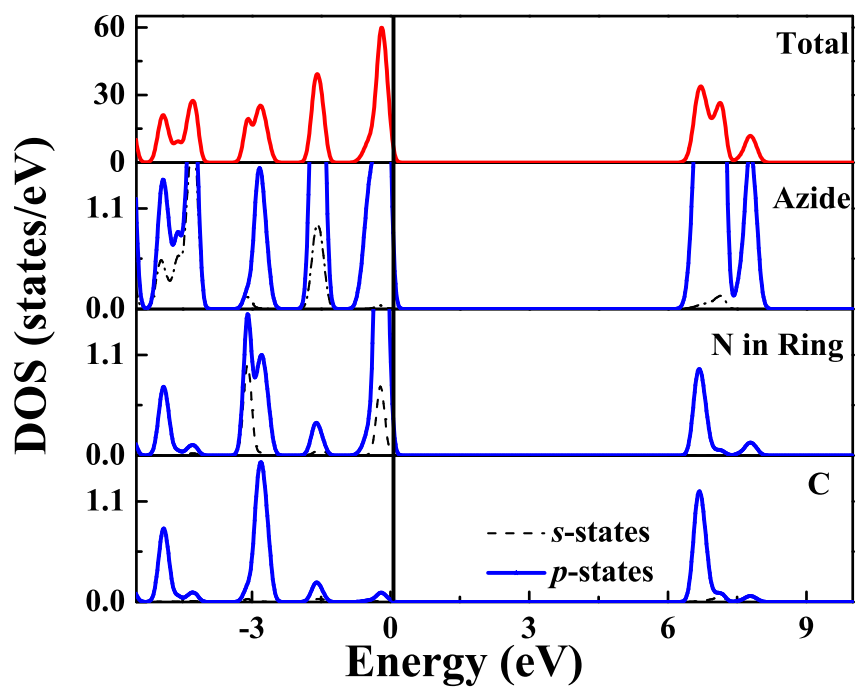}}
\caption{(Colour online) Total and partial density of states (DOS) of C$_3$N$_3$(N$_3$)$_3$ }
\end{center}
\end{figure}
\clearpage
\newpage

\begin{figure}
\begin{center}
{\includegraphics[width=6in,clip]{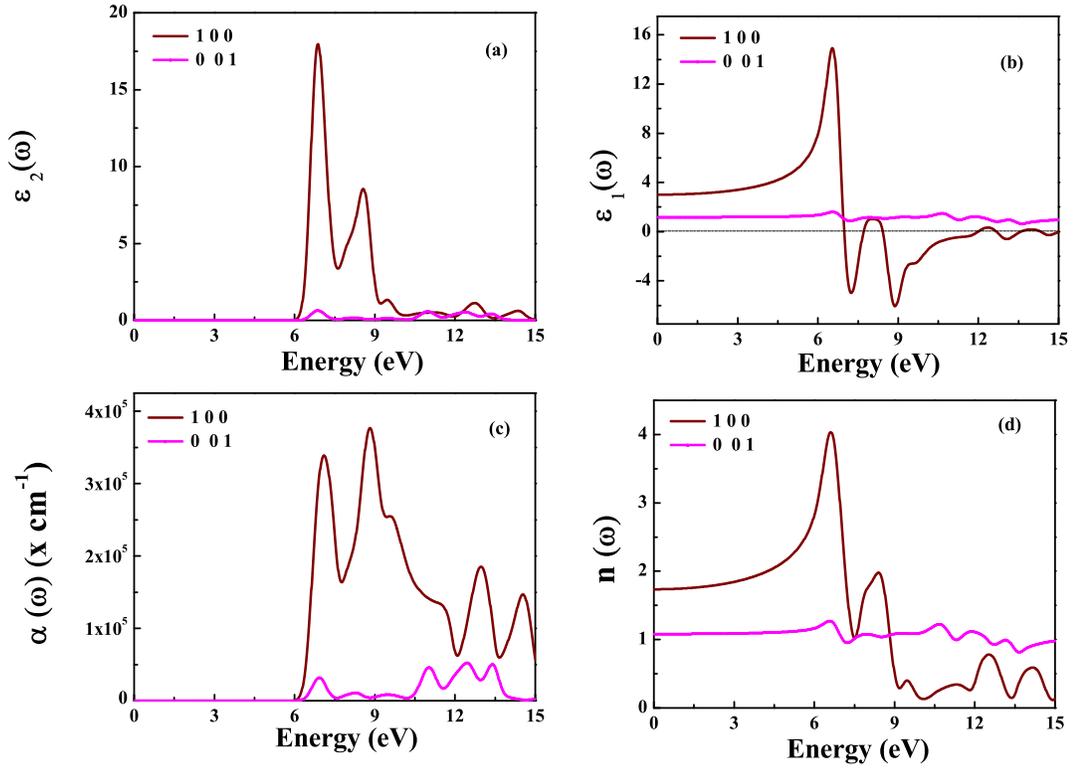}}
\caption{(Colour online) Optical properties , (a) absorptive part of the dielectric function (b)  dispersive part of the dielectric function (c) refractive index and (d) absorption of C$_3$N$_3$(N$_3$)$_3$ as calculated at theoretical equilibrium with a scissor shift of 2.68 eV. }
\end{center}
\end{figure}
\clearpage
\newpage

\begin{table}[ht]
\caption{ The calculated lattice parameters  (a, c in \AA) and volume (V in \AA{$^3$}) of C$_3$N$_3$(N$_3$)$_3$ by using three DFT methods LDA, PBE, and PBE+G06. Experimental data have been taken from Elmer et. al \cite{Elmer}}
\begin{tabular}{cccccccccccccccccccccccccccccc} \hline
 axis     &LDA      &PBE        & PBE+G06      &   Exp   \\ \hline \hline
 a        & 8.74    & 9.74      & 9.00         & 8.75    \\
 c        & 5.40    & 6.42      & 5.70         & 5.89    \\
 V        & 357.06  & 527.32    & 399.39       & 390.44  \\\hline \hline
\end{tabular}
\end{table}

\begin{table}[ht]
\caption{ Single crystal elastic constants (C$_{ij}$, in GPa) and bulk modulus (B in GPa)  of C$_3$N$_3$(N$_3$)$_3$ at the respective theoretical equilibrium volume obtained with the PBE+G06 functional.}
\begin{tabular}{cccccccccccccccccccccccccccccc} \hline
 C$_{11}$ & C$_{33}$ & C$_{44}$ & C$_{12}$ & C$_{13}$ & B \\ \hline \hline
 25.08    & 16.53    & 3.63     & 3.55     & 5.65     & 10.45 \\ \hline
\end{tabular}
\end{table}

\end{document}